\documentclass[onecolumn,showpacs,preprintnumbers,amsmath,amssymb, longbibliography]{revtex4}
\usepackage{graphicx}
\usepackage{dcolumn}
\usepackage{bm}
\usepackage{epsfig}
\usepackage{subfigure}
\usepackage{xcolor}
\usepackage{bbold}
\usepackage{hyperref}

\newcommand{\bi}{\begin{itemize}}
\newcommand{\ei}{\end{itemize}}
\newcommand{\be}{\begin{eqnarray}}
\newcommand{\ee}{\end{eqnarray}}
\newcommand{\beq}{\begin{equation}}
\newcommand{\eeq}{\end{equation}} 

\newcommand{\besub}{ \begin{subequations} }
\newcommand{\eesub}{ \end{subequations} }

\newcommand{\bbmatrix}{\left( \begin{array}}
\newcommand{\eematrix}{\end{array} \right)}

\newcommand{\dd}{\text{d}}

\newcommand{\p}{\partial}
\newcommand{\norm}[1]{\left\lVert#1\right\rVert}

\def\ch#1{\textcolor{black}{#1}}

\begin{document}

\title{Stochastic optimal control formalism for an open quantum system}

\author{Chungwei Lin$^1$\footnote{clin@merl.com}, Dries Sels$^{2,3,4}$, Yanting Ma$^1$, Yebin Wang$^1$}
\date{\today}
\affiliation{$^1$Mitsubishi Electric Research Laboratories, 201 Broadway, Cambridge, MA 02139, USA \\ 
$^2$Department of physics, Harvard University, Cambridge, MA 02138, USA \\ 
$^3$Department of physics, New York University, New York City, NY 10003, USA \\
$^4$Center for Computational Quantum Physics, Flatiron Institute, 162 5th Ave, New York, NY 10010, USA
}

\begin{abstract} 
A stochastic procedure is developed which allows one to express Pontryagin's maximum principle for dissipative quantum system solely in terms of stochastic wave functions. Time-optimal controls can be efficiently computed without computing the density matrix. Specifically, the proper dynamical update rules are presented for the stochastic costate variables introduced by Pontryagin's maximum principle and restrictions on the form of the terminal cost function are discussed. The proposed procedure is confirmed by comparing the results to those obtained from optimal control on  Lindbladian dynamics. Numerically, the proposed formalism becomes time and memory efficient for large systems, and it can be generalized to describe non-Markovian dynamics.
\end{abstract}

\maketitle

\section{Introduction}

Modern quantum technology \cite{Georgescu_2012}, including quantum computation \cite{Neilson_book, book:Kaye_book, Shor:1997:PAP:264393.264406, Grover:1996:FQM:237814.237866, PhysRevLett.79.325, Peruzzo-2014, Farhi_14, PhysRevX.6.031007, Preskill2018quantumcomputingin}, quantum sensing / metrology / imaging \cite{book:Helstrom, book:Holevo, PhysRevLett.96.010401, AdvancesQuantumMetrology_2011, PhysRevX.6.031033, PhysRevA.96.040304, PhysRevLett.117.110801, LIGO_2011, doi:10.1116/1.5119961, RevModPhys.90.035005}, and quantum communication \cite{PhysRevLett.69.2881, PhysRevA.61.042302, PhysRevLett.67.661, 144km, RevModPhys.66.481, RevModPhys.77.513, RevModPhys.84.621},  commonly rely on coherent control of the state of the system. A typical quantum task starts from an easily prepared initial state, undergoes a designed control protocol, and hopefully ends up with a state sufficiently close to the target state (not necessarily known in advance). When the closeness to the target state can be characterized by a scalar ``terminal cost function'', this quantum problem can be mathematically formulated as an optimal control problem. Many quantum applications (or at least an intermediate step of the application) fit this description. Relevant examples include the state preparation \cite{PhysRevA.97.062343, ahmed19, PhysRevX.8.021012, PhysRevLett.112.047601} where the cost function is the overlap to the known target state, the ``continuous-time'' variation-principle based quantum computation \cite{Farhi_00, PhysRevLett.103.080502, PhysRevA.90.052317, Peruzzo-2014, PhysRevX.6.031007} where the cost function is the ground state energy, and the quantum parameter estimation \cite{RevModPhys.90.035005, PhysRevLett.124.060402} where the cost function is the quantum Fisher information.

Pontryagin's Maximum Principle (PMP) \cite{book:Pontryagin, Sussmann_87_01, book:Luenberger, book:GeometricOptimalControl} is a powerful formalism in classical control theory, and it 
has been applied to quantum state preparation \cite{PhysRevA.97.062343, PhysRevA.101.022320}, non-adiabatic quantum computation \cite{PhysRevX.7.021027, PhysRevA.100.022327}. Due to the linearity of Schr\"odinger's equation, PMP implies the time-optimal control typically has the so called \emph{bang-bang} form (the control takes its extreme values); the bang-bang form puts strong constraints on the structure of optimal solutions and is found to be the case in some problems \cite{PhysRevA.97.062343, PhysRevA.98.012301, PhysRevX.7.021027}. For general quantum problems, however, the optimal control often includes a singular part \cite{PhysRevLett.111.260501, PhysRevX.8.031086, PhysRevA.100.022327, PhysRevA.101.022320}, which makes PMP less informative about the solution.
While one can in principle explicitly solve for the behavior one the singular arcs~\cite{PhysRevA.101.022320}, such analysis is restricted to small systems as it quickly becomes intractable.  The main usefulness of PMP thus appears to be numerical: first, it provides an efficient way to compute the gradient of the terminal cost function; second, PMP gives the necessary conditions for an optimal solution which can be used to check the quality of any numerical solutions.

The application of PMP to quantum dissipative systems using a Lindblad master equation approach  has been proposed and discussed in literatures \cite{PhysRevA.76.023419, PhysRevA.78.052112, PhysRevA.80.045401, Ritland_2018}. 
In Ref.~\cite{PhysRevA.101.022320} we demonstrated that the singular controls are essential in open systems by examining the PMP optimality conditions.
In this paper, we develop a stochastic formalism to evaluate important quantities introduced by PMP. Specifically we derive a procedure to consistently update the wave function and its costate for the stochastic Schr\"odinger equation so that the optimal control can be determined without constructing the density matrices. The procedure is checked against the results obtained using the Lindbladian equation.
The proposed stochastic procedure not only saves time and memory for simulating large systems, but can be helpful for problems where the wave function description is intuitive, such as gate-based quantum computation \cite{Deutsch_Jozsa_92, Simon:1997:PQC:264393.264405, Shor:1997:PAP:264393.264406, Grover:1996:FQM:237814.237866, PhysRevLett.79.325, Bennett_97, PhysRevLett.103.150502}, quantum error correction \cite{PhysRevA.52.R2493, PhysRevLett.77.793, PhysRevA.55.900, Gottesman-1997} and coherent feedback control  \cite{Hirose-2016}. Moreover, the generalization to describe systems coupled to a non-Markovian bath \cite{PhysRevA.58.2733, PhysRevLett.82.2417, PhysRevA.87.052328} is straightforward.

The rest of the paper is organized as follows. In Section~\ref{sec:PMP_DM} we  recapitulate relevant conclusions from classical control theory. In particular, we express important quantities introduced by PMP in terms of density matrix which are essential for open quantum systems. 
In Section~\ref{sec:PMP_stoch} we describe how to use the stochastic Schr\"odinger equation to simulate the dissipative system, with an emphasis on determining the optimal control. 
In Section~\ref{sec:numerics} we numerically implement the proposed stochastic procedure to the single qubit system. In particular we compare the results from the deterministic Lindbladian with those from the stochastic Schr\"odinger equation. 
A brief conclusion is given in Section~\ref{sec:Conclusion}.

\section{Density matrix formulation and Pontryagin's maximum principle} 
\label{sec:PMP_DM}

Consider a quantum system characterized by the density matrix $\rho$, evolving in time under some general Markovian quantum dynamics 
\begin{equation}
\dot{\rho}=\mathcal{L}_t[\rho],
\end{equation}
with a $\mathcal{L}_t$ a Liouvillian superoperator which depends on some control variable $u(t)$. In particular, we will be concerned with dynamical systems with linear controls such that  $\mathcal{L}_t=\mathcal{L}_0+u(t) \mathcal{L}_u$, where $\mathcal{L}_0$ represents the bare dynamics of the system beyond our control and $\mathcal{L}_u$ is the controllable part. The control field $u(t)$ is further assumed to be bounded $|u(t)| \leq 1$. While most of our discussion is completely general, and applies to arbitrary Liouvillians, we are primarily concerned with coherent controls, i.e., 
\begin{equation}
\mathcal{L}_u[\rho]= -i[H_u,\rho],
\end{equation}
where $H_u$ is the part of the Hamiltonian that can be controlled. Moreover, the present goal is to maximize the overlap between the final state at time $t_f$ and the target state $\left| \psi_{\rm tar} \right>$, which can be expressed by the terminal cost function $\mathcal{C}$ (to minimize)
\beq 
\mathcal{C}( t_f) = -  \text{Tr} [ \rho(t_f) \rho_\text{tar} ] 
= - \langle \psi_\text{tar} | \rho(t_f) | \psi_\text{tar} \rangle.
\label{eqn:cost_QFidelity}
\eeq  
This cost function has the benefit of being linear in the state $\rho(t_f)$. \ch{Notice that for a pure target state $\langle \psi_\text{tar} | \rho(t_f) | \psi_\text{tar} \rangle$  corresponds to the standard Uhlmann fidelity \cite{UHLMANN1976273, doi:10.1080/09500349414552171}. Here we shall focus the discussions to a target state that is pure but since the cost function is just the Hilbert-Schmidt inner product between the terminal state and the target state, the same cost function could be used to mixed states as well.}  Following PMP, we proceed by introducing the control Hamiltonian (c-Hamiltonian) $\mathcal{H}_c$
\beq 
\begin{aligned}
\mathcal{H}_c &=  \text{Tr}\left[ 
\lambda \dot{\rho} \right] =   \text{Tr}\left[ 
\lambda \mathcal{L}_0[\rho] \right] + u(t) \,
\text{Tr}\left[ \lambda \mathcal{L}_u[\rho] \right],
\end{aligned}
\label{eqn:Hc_basic}
\eeq
which is a {\em real-valued} scalar and should not be confused with the Hamiltonian $H$ of the system \cite{Hc_basic2}. The c-Hamiltonian $\mathcal{H}_c$ is constructed such that one of Hamilton's equations simply yields the equation of motion for the state, i.e.,
\begin{equation}
\dot{\rho}= \frac{\partial \mathcal{H}_c}{\partial \lambda},
\label{eq:state}
\end{equation}
where the canonical momentum $\lambda$, typically referred to as costate in the context of optimal control, satisfies 
\begin{equation}
\dot{\lambda}= - \frac{\partial \mathcal{H}_c}{\partial \rho}. 
\label{eq:co_sigma}
\end{equation}
The initial condition for the state is typically supplied $\rho(t_0)=\rho$ and according to PMP the boundary condition for the costate should satisfy
\begin{equation}
\lambda(t_f)= \frac{\partial \mathcal{C}(t_f)}{\partial \rho(t_f)}= - \left| \psi_{\rm tar} \right> \left<  \psi_{\rm tar}  \right|
\end{equation}
Necessary conditions for optimal solutions to the time-optimal control problem are to simultaneously satisfy Eqs.~\eqref{eq:state},~\eqref{eq:co_sigma}, together with $\mathcal{H}_c= {\rm constant}$ over the entire interval $t_0$ to $t_f$ 
and 
\begin{equation}
u(t) = \begin{cases} +1 & \text{ if } \Phi(t)<0 \\
          -1 & \text{ if } \Phi(t)>0      \\
          \text{undetermined} & \text{ if } \Phi(t)=0 \end{cases},
\label{eqn:bang_condition}
\end{equation}
with the switching function $\Phi(t)= \text{Re} \left( \text{Tr}\left[ 
\lambda \mathcal{L}_u[\rho] \right] 
\right)$, which for coherent controls becomes
\begin{equation}
\Phi(t)= \text{Im} \left( \text{Tr}\left[ 
\lambda [H_u,\rho] \right] 
\right). 
\label{eq:switch}
\end{equation}
It is worth noting that the switching function corresponds to the gradient of the terminal cost function, i.e., $\Phi(t) \sim \frac{\p \mathcal{C} }{\p u(t)}$. When the goal is to minimize $\mathcal{C}$, $u(t)$ takes the extreme value with a sign opposite to $\Phi(t) \neq 0$; this is referred to as ``bang control'' and is exactly Eq.~\eqref{eqn:bang_condition}. When $\Phi(t)=0$ for a finite amount of time the controls might not be extremal and their structure can be determined by looking at higher time derivatives of the switching function. This is beyond the present discussion but some complementary derivations, including a discussion of singular controls can be found in Refs.~\cite{PhysRevA.100.022327, PhysRevA.101.022320}. 

Before moving on to discuss stochastic dynamics, let us consider some particularly relevant dynamics. First of all, consider the dynamics to be unitary such that $\mathcal{L}[\rho]=-i[H,\rho]$. The c-Hamiltonian then reads
\begin{equation}
\mathcal{H}_c = \text{Im} \left( \text{Tr}\left[ 
\lambda [H,\rho] \right] \right)= -\text{Im} \left( \text{Tr}\left[ 
\rho [H,\lambda] \right] \right), 
\label{eqn:DM_costate_Hoc}
\end{equation}
where the latter just follows from the cyclic properties of the trace. Consequently the costate evolution,
\begin{equation}
\dot{\lambda}=-i [H,\lambda],
\end{equation}
is identical to that of the state $\rho$. This is not the case for dissipative dynamics, consider for example 
\begin{equation}
\mathcal{L}[\rho]=L \rho L^\dagger - \frac{1}{2}  \left( L^\dagger L \rho  +  
\rho L^\dagger L \right),
\label{eqn:Lindblad}
\end{equation}
where $L$ are Lindblad jump operators.  The c-Hamiltonian now takes the form 
\begin{eqnarray}
\mathcal{H}_c &=& \text{Re} \left( \text{Tr}\left[ 
\lambda \left\lbrace L \rho L^\dagger - \frac{1}{2}  \left( L^\dagger L \rho  +  
\rho L^\dagger L \right) \right\rbrace  \right] \right) \nonumber \\
&=& \text{Re} \left( \text{Tr}\left[ 
\rho \left\lbrace L^\dagger \lambda L - \frac{1}{2}  \left( L^\dagger L \lambda  +  
\lambda L^\dagger L \right) \right\rbrace  \right] \right),
\label{eqn:DM_costate_Hoc_diss}
\end{eqnarray}
cosequently the costate evolution, 
\begin{equation}
\dot{\lambda}=  -\left[ L^\dagger \lambda L - \frac{1}{2}  \{ L^\dagger L, \lambda \} \right],
\label{eqn:lambda_Lindblad}
\end{equation}
where the anticommitator is defined as $\{A, B\} = AB+BA$. Note that in the case of Hermitean jump operators Eq.~\eqref{eqn:lambda_Lindblad} simply amounts to a time reversal $t \rightarrow -t$ of Eq.~\eqref{eqn:Lindblad}. The sum of Eq.~\eqref{eqn:DM_costate_Hoc} and \eqref{eqn:DM_costate_Hoc_diss} gives the c-Hamiltonian for systems having both unitary and dissipative dynamics.
In principle, optimal control solutions can be found by iteratively solving for the (co)-state, extracting the switching function and updating the controls \cite{PhysRevA.101.022320}. The downside is that one has to explicitly propagate
 the density matrix and its costates.

\section{Stochastic control}
\label{sec:PMP_stoch}

In this section we describe the proposed stochastic procedure that extracts the switching function and related quantities from a stochastic Schr\"odinger simulation by properly correlating the Poisson random processes between the wave function and its costate. The limitation of the procedure will also be pointed out.

\subsection{Stochastic simulation of density matrix} 

The formalism to simulate the density matrix $\rho$  [Eq.~\eqref{eqn:Lindblad}] by averaging stochastic wave function has been developed in the early days of quantum optics \cite{book:open_quantum, PhysRevLett.68.580, castin2008wave}. To facilitate the later discussion, we briefly review it.
Denoting $|\dd  \psi \rangle \equiv | \psi (t + \dd t) \rangle - | \psi(t) \rangle$, the stochastic update of the wave function is given by 
\besub
\begin{align}
|\dd  \psi \rangle &=  G | \psi \rangle  \, \dd t + 
\left(  L | \psi \rangle  - | \psi \rangle \right) \dd N(t), 
\label{eqn:psi_random}\\
\text{where }& G  | \psi \rangle 
= \left[ -iH - \frac{\gamma}{2} \,  L^\dagger L
+  \frac{\gamma}{2}  \mathbb{1}  \right] | \psi \rangle.
\label{eqn:psi_nonUni}
\end{align}
\label{eqn:Sto_Poisson_A0}
\eesub
In Eqs.~\eqref{eqn:Sto_Poisson_A0}, $\dd N(t)$ is the Poisson random variable whose mean and variance are both $\gamma \, \dd t$, i.e., $\dd N(t) = \dd N(t)^2 = \gamma \, \dd t$; $H$ is the Hamiltonian of the system. 
The wave function evolves forwardly in time as 
$ | \psi (t) \rangle \rightarrow 
| \psi (t + \dd t) \rangle = | \psi (t) \rangle + | \dd \psi (t) \rangle$.  
The density matrix can then simply be extracted as $\rho= \mathbb{E} [\left| \psi(t) \right> \left< \psi(t) \right| ]$, with the expectation value taken over the Poisson process (with the same initial state $| \psi_\text{ini} \rangle$) that generates the quantum jumps: 
\beq 
\rho(t) = \mathbb{E} [\left| \psi(t) \right> \left< \psi(t) \right| ] \approx \frac{1}{N} \sum_{n=1}^N | \psi^{(n)}(t) \rangle
\langle \psi^{(n)} (t) |,
\label{eqn:DM_stochastic}
\eeq 
where the superscript $n$ labels $n$th realization.
The same procedure can be repeated for the costate $\lambda$ by unraveling it in stochastic wave function $\left| \pi \right>$. Recall that the costate needs to evolve backwards in time from $t_f$, hence denoting $|-\dd  \pi \rangle \equiv | \pi (t - \dd t) \rangle - | \pi(t) \rangle$, one has
\besub
\begin{align}
|-\dd \pi \rangle &= \tilde{G}  | \pi \rangle  \, \dd t + 
\left(  L^{ {\dagger} } | \pi \rangle  - | \pi \rangle \right) \dd N(t), 
\label{eqn:pi_random} \\
\text{and }& \tilde{G}  | \pi \rangle  
= \left[ iH - \frac{\gamma}{2}   L^\dagger L  
+  \frac{\gamma}{2}  \mathbb{1}  \right] | \pi \rangle.
 \label{eqn:pi_nonUni}
\end{align}
\label{eqn:Sto_Poisson_pi_A0}
\eesub
$\dd N(t)$ is again the Poisson random variable with mean and variance of $\gamma \, \dd t$; futher note that $\tilde{G}=G^\dagger$.  
To extract $\lambda$, one repeats Eq.~\eqref{eqn:Sto_Poisson_pi_A0} to obtain $| \pi^{(n)} \rangle$ starting the same final state $| \pi (t_f) \rangle$, the costate density matrix is then computed using  
\beq 
\lambda(t) =\mathbb{E} [\left| \pi(t) \right> \left< \pi(t) \right| ] \approx \frac{1}{N} \sum_{n=1}^N | \pi^{(n)}(t) \rangle
\langle \pi^{(n)} (t) |.
\label{eqn:lambda_stochastic}
\eeq   

It is worth noting that the stochastic procedure outlined in Eqs.\eqref{eqn:DM_stochastic} and \eqref{eqn:lambda_stochastic} does {\em not} apply to cases of negative $\gamma$ despite the Lindbladian equation can be solved no matter the sign of $\gamma$. Since the stochastic procedure always increases the entropy, it can only describe the forward propagation of $\rho$ and backward propagation of $\lambda$.  We also point out that both Eq.~\eqref{eqn:DM_stochastic} and \eqref{eqn:lambda_stochastic} can be used to simulate the mixed state. One simply has to, in conjunction with sampling the Poisson process, take random samples out of the initial density matrix. 

 
\subsection{Two stochastic procedures for switching function \label{sec:two_procedures}}  

For the specific cost function given by expression~\eqref{eqn:cost_QFidelity}, the boundary condition for the costate becomes $\lambda (t_f) = - \left| \psi_{\rm tar} \right> \left< \psi_{\rm tar} \right|$. The switching function $\Phi(t)$ can thus straightforwardly be computed from  Eq.~\eqref{eq:switch} with $\rho(t)$ from Eq.~\eqref{eqn:DM_stochastic} and $\lambda(t)$ from Eq.~\eqref{eqn:lambda_stochastic}.
Once $\rho(t)$ and $\lambda(t)$ are known, the c-Hamiltonian can also be evaluated using Eq.~\eqref{eqn:DM_costate_Hoc} and \eqref{eqn:DM_costate_Hoc_diss}. This procedure works generally and a numerical example will be provided in Section \ref{sec:example_phi}.
In this procedure, $\rho$ and $\lambda$ have to be constructed explicitly which makes the procedure numerically quite involved, e.g., one has to store all stochastic realizations of $|\psi^{(n)}(t) \rangle$ and $|\lambda^{(n)}(t) \rangle$, and then explicitly perform the trace in~\eqref{eq:switch}. 

In the first procedure just described, the Poisson random variables that generate the state and the costate are completely uncorrelated. By correlating the random variables, however, the switching function can be obtained without computing $\rho$ and $\lambda$ explicitly as long as the terminal cost function is a linear function of the state $\rho$. This is natural in many situations and the present cost function~\eqref{eqn:cost_QFidelity} is clearly of that form.

To see how the second procedure works, consider first the cost function~\eqref{eqn:cost_QFidelity}. Expressed in terms of stochastic wave functions $\left| \psi \right>$ this becomes 
\begin{equation}
\mathcal{C}(t_f) = - \left< \psi_\text{tar} \right| \mathbb{E}\left[ \left| \psi \right> \left< \psi \right| \right] \left| \psi_{\rm tar} \right> = - \mathbb{E} \left[ \vert \left< \psi_{\rm tar} | \psi \right> \vert^2\right].
\end{equation}
The linearity assures that we can interchange the order in which we take the quantum expectation value and the average of the classical random process. Hence, consider the $n$th stochastic realization defined by the Poisson random process $\dd N^{(n)}(t)$, the $| \psi^{(n)} \rangle$ and its costate $| \pi^{(n)} \rangle$, where the latter satisfy Eqs.~\eqref{eqn:psi_random} and~\eqref{eqn:pi_random} (with the same $dN(t)$) respectively, then 
\begin{equation}
\mathcal{C}(t_f) \approx -\frac{1}{N} \sum_n \left\vert \left< \psi_{\rm tar} | \psi^{(n)} (t_f) \right> \right\vert^2 \equiv -\frac{1}{N} \sum_n \mathcal{C}^{(n)}(t_f) .
\label{eqn:cost_stoch}
\end{equation}
Consequently, according the PMP, the boundary condition for $n$th realization of the costate $|\pi^{(n)}(t) \rangle$ is fixed by
\beq 
|\pi^{(n)}(t_f) \rangle = 
\frac{\p\, \mathcal{C}^{(n)}(t_f)}{ \left< \psi^{(n)}(t_f) \right|} =
- \, |  \psi_\text{tar} \rangle \langle \psi_\text{tar} |  \psi^{(n)}(t_f) \rangle . 
\label{eqn:stochastic_pi_BC}
\eeq 
The switching function is then computed 
\beq 
\begin{aligned} 
\Phi(t) = \frac{1}{N} \sum_{n=1}^N \Phi^{(n)}(t), \text{ with }
\Phi^{(n)}(t) =\text{Im} \langle \pi^{(n)}(t) |  H_u  | \psi^{(n)}(t) \rangle.
\end{aligned} 
\label{eqn:quantum_Hoc_switch_2_2}
\eeq 
In contrast to the naive (first) procedure, the latter does not explicitly estimate any density matrix which saves time and computer memory. Note that, apart from the Poisson random processes used in $| \psi \rangle$ and $ | \pi \rangle$ being identical in very realization, the $n^{\rm th}$ costate also explicitly depends on the $n^{\rm th}$ solution of the state $|\psi \rangle$ through its boundary condition. 

Eq.~\eqref{eqn:quantum_Hoc_switch_2_2} can be generalized to compute other quantities. In particular, Im($\text{Tr} \left[ \lambda [H, \rho] \right]$) in Eq.~\eqref{eqn:DM_costate_Hoc} can be evaluated by replacing $H_u$ by $H(t)$ in Eq.~\eqref{eqn:quantum_Hoc_switch_2_2};  the anti-commutator Re($\text{Tr} \lambda \{ L^\dagger L, \rho \} $) in  Eq.~\eqref{eqn:DM_costate_Hoc_diss} can be evaluated by averaging over $\text{Re}( \langle \pi^{(n)}(t) | L^\dagger L  | \psi^{(n)}(t) \rangle )$. These expressions are numerically tested (not shown). 
The c-Hamiltonian $\mathcal{H}_c$, however, can only be computed  using the first stochastic procedure because $\text{Tr} (\lambda L \rho L^\dagger )$ cannot be expressed as the average of stochastic realizations in the form of Eq.~\eqref{eqn:quantum_Hoc_switch_2_2}.

To conclude this section, we notice that Eq.~\eqref{eqn:Sto_Poisson_A0} is by no means the unique stochastic unraveling. 
Notably when $L=L^\dagger$, one can define $H-i \gamma(t) L$ as the non-Hermitean Hamiltonian where $\gamma(t)$ is the Gaussian random process with a $\dd t$-dependent variance; averaging many stochastic realizations also properly simulates the Lindblad equation \cite{PhysRevLett.68.580, castin2008wave}. 
Both proposed procedures for switching function, particular the second one that correlates the random variables for $|\psi \rangle$ and $|\pi \rangle$, apply to this stochastic implementation as well (tested, not shown). 
An interesting open question is to consider non-Markovian random process, beyond the Lindbladian formalism. 



\section{Numerical implementation} 
\label{sec:numerics}
In this section we confirm the proposed formalism by applying it to the specific single qubit problem where the numerically exact solutions are non-trivial but known. 

\subsection{Single qubit problem}  

\begin{figure}[ht]
\begin{center}
\includegraphics[width=0.5\textwidth]{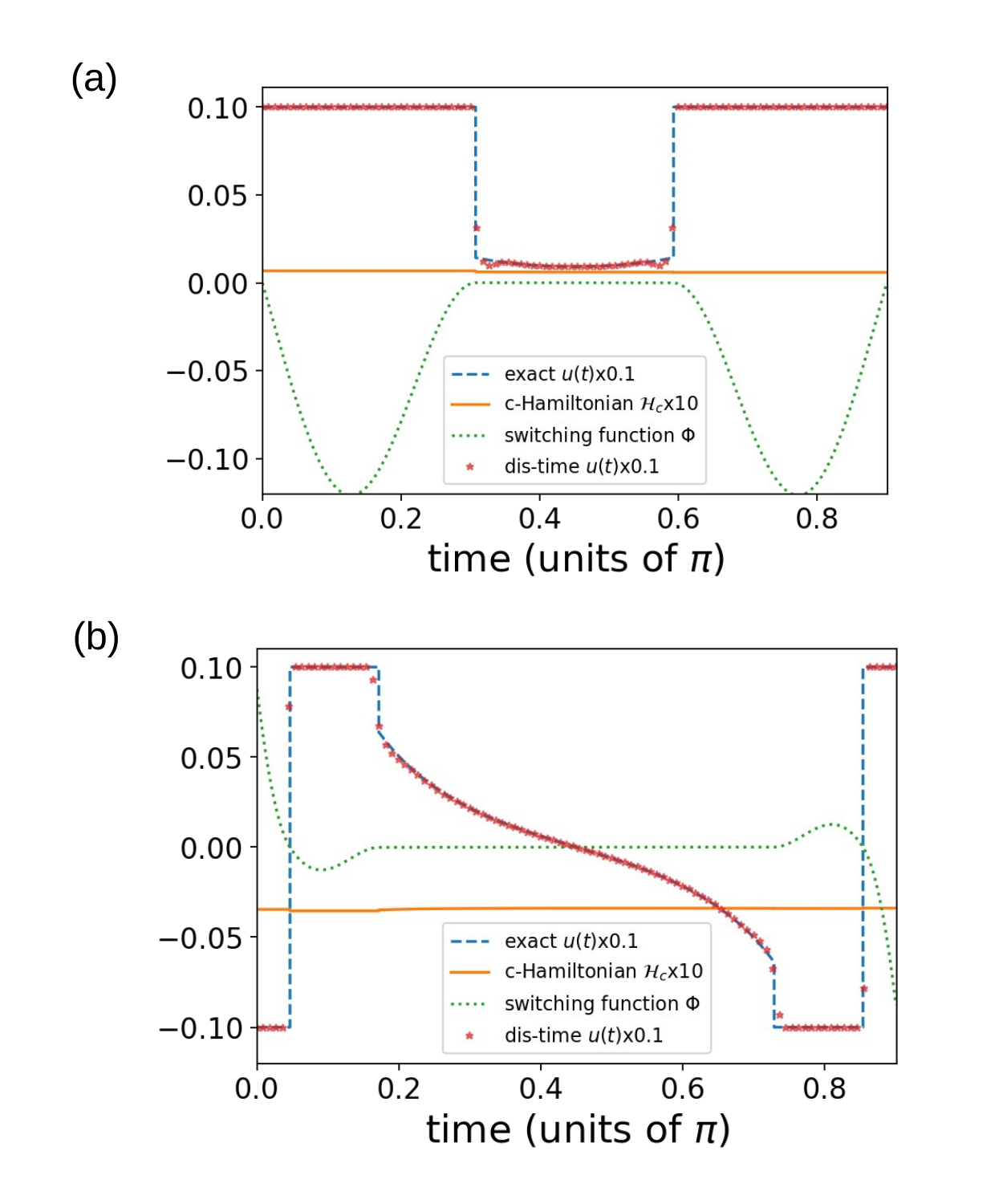}
\caption{The optimal controls (dashed curves) for (a) state retention  and (b) state preparation  problem. A bang control corresponds to non-zero $\Phi$ (dotted curves) whereas a singular control to the vanishing $\Phi$. 
The exact optimal controls (dashed curves) are obtained using the formalism developed in Ref.~\cite{PhysRevA.101.022320}. 
The c-Hamiltonian (solid curves) is a constant over the entire evolution time $t_f$.  The dotted curve are the numerical solutions obtained using gradient-based method with the switching function computed from solving the Lindbladian equation; a good agreement is seen.
}
\label{fig:reference}
\end{center}
\end{figure} 

To numerically test the proposed formalism, we consider a dissipative qubit system where the density matrix satisfies the Lindbladian equation:  
\beq 
\begin{aligned} 
\frac{\p }{\p t}\rho &= -i [H(t), \rho] +  \gamma \left[ L \rho L^\dagger- \frac{1}{2} \{ L^\dagger L, \rho \} \right]
\end{aligned}
\label{eqn:Problem}
\eeq 
For unitary dynamics, we consider the Landau-Zener type Hamiltonian where $H(t) = H_0 + u(t) H_u = \sigma_x  + u(t)  \sigma_z$ with $u(t)$ being the single control field bounded by $|u(t)| \leq 1$ and $\sigma$'s denoting Pauli matrices \cite{PhysRevLett.111.260501, PhysRevX.8.031086}. For dissipative dynamics, we choose $L = \sigma_x$, $\gamma = 0.5$, and a total evolution time of $t_f = 0.9 \pi$. These parameters produce non-trivial control protocols  \cite{PhysRevA.101.022320}; in particular, the combined choice of $H$ and $L$ leads to an optimal control that prevents the system from decaying to the maximal-entropy state even when $t_f \rightarrow \infty$, independent of the initial and target states. 

Two sets of initial and target states are considered: the state retention problem where the initial and target states $| \psi_\text{ini} \rangle$ and  $| \psi_\text{tar} \rangle$ are
\beq 
| \psi_\text{ini} \rangle = | \psi_\text{tar} \rangle = \begin{bmatrix} 1\\ 0 \end{bmatrix};
\label{eqn:state_retention}
\eeq 
and the state preparation problem where
\beq 
\begin{aligned}
|\psi_\text{ini} \rangle &= \frac{1}{ \sqrt{ 10+4\sqrt{5} }  } \begin{bmatrix} 1\\ -2-\sqrt{5} \end{bmatrix}, \\
|\psi_\text{tar} \rangle &= \frac{1}{ \sqrt{ 10-4\sqrt{5} }  } \begin{bmatrix} 1 \\ 2-\sqrt{5}  \end{bmatrix}. 
\end{aligned}
\label{eqn:state_preparation}
\eeq 
The corresponding density matrix is given by $\rho = | \psi \rangle \langle \psi|$: for the state retention problem, $\rho_\text{ini} = \rho_\text{tar} = \frac{1}{2} ( \mathbb{I} + \sigma_x)$; for the preparation, $\rho_\text{ini (tar)} = \frac{1}{2} \left[ \mathbb{I} - \frac{1}{\sqrt{5}} \sigma_x - (+) \frac{2}{\sqrt{5}} \sigma_z \right]$. The boundary condition of costate is given by $\lambda(t_f) = -| \psi_\text{tar} \rangle \langle \psi_\text{tar} |$. The choices of Eq.~\eqref{eqn:state_retention} and \eqref{eqn:state_preparation} describe two limits: the initial and target states are close to (identical here) or far away from each other. Using the formalism developed in Ref.~\cite{PhysRevA.101.022320}, the optimal controls, switching functions, and c-Hamiltonians for both problems are given in Fig.~\ref{fig:reference}; these solutions are referred to as the ``exact'' solutions and will be served as the reference for comparison.

\subsection{Switching functions}
\label{sec:example_phi}

\begin{figure}[ht]
\begin{center}
\includegraphics[width=0.5\textwidth]{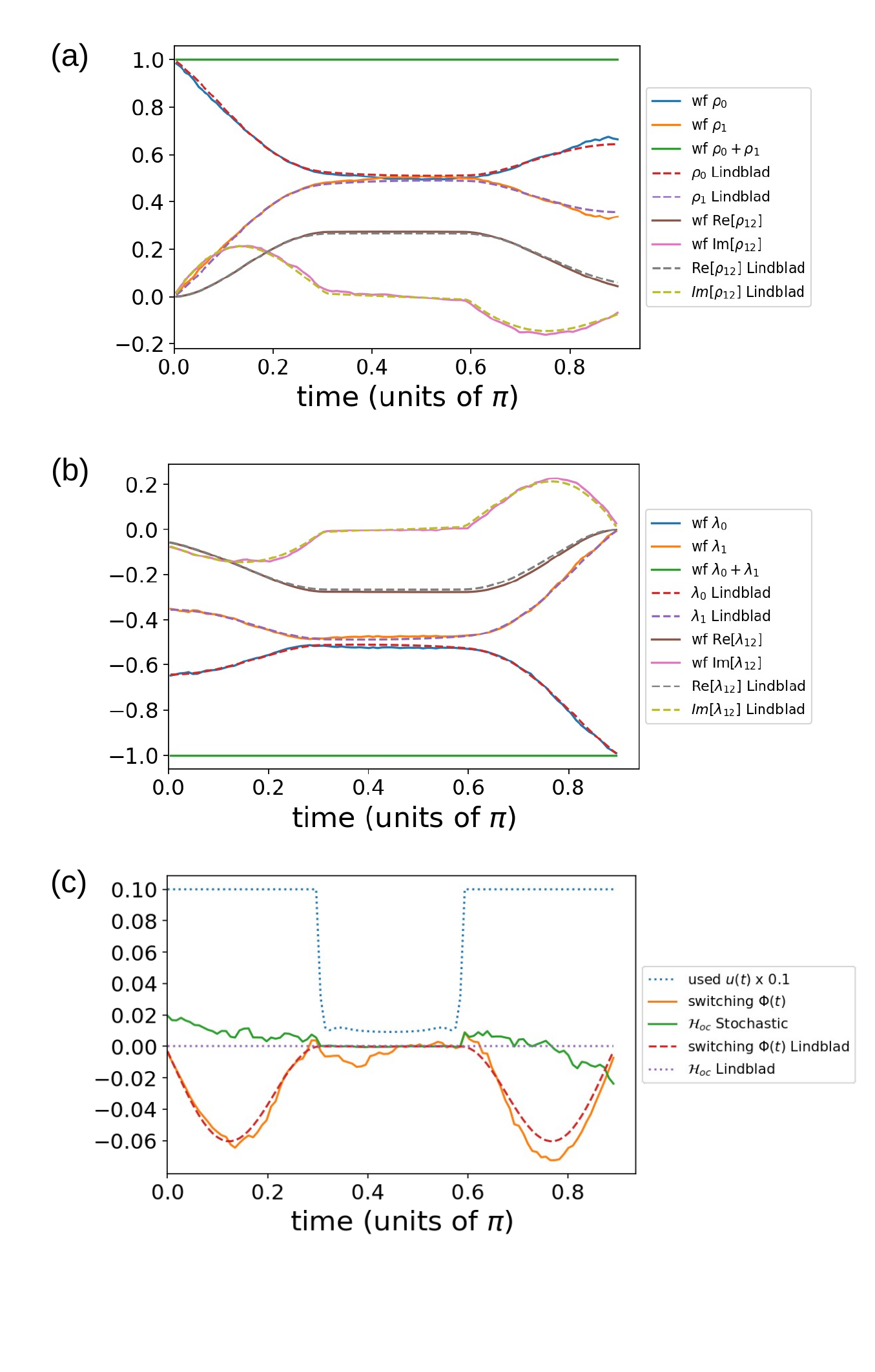}
\caption{Comparison between the deterministic Lindbladian (dashed curves) and the stochastic Schr\"odinger (solid curves) simulations for (a) density matrix $\rho$ and (b) the costate density matrix $\lambda$. The control is chosen to be the optimal control for the state retention problem (dotted curve in (c). The stochastic results use 100 time points and average over 500 realizations. (c) Combining $\rho$ and $\lambda$ to evaluate the switching function and c-Hamiltonian.
}
\label{fig:WF_DM_comparison}
\end{center}
\end{figure}  

\begin{figure}[ht]
\begin{center}
\includegraphics[width=0.5\textwidth]{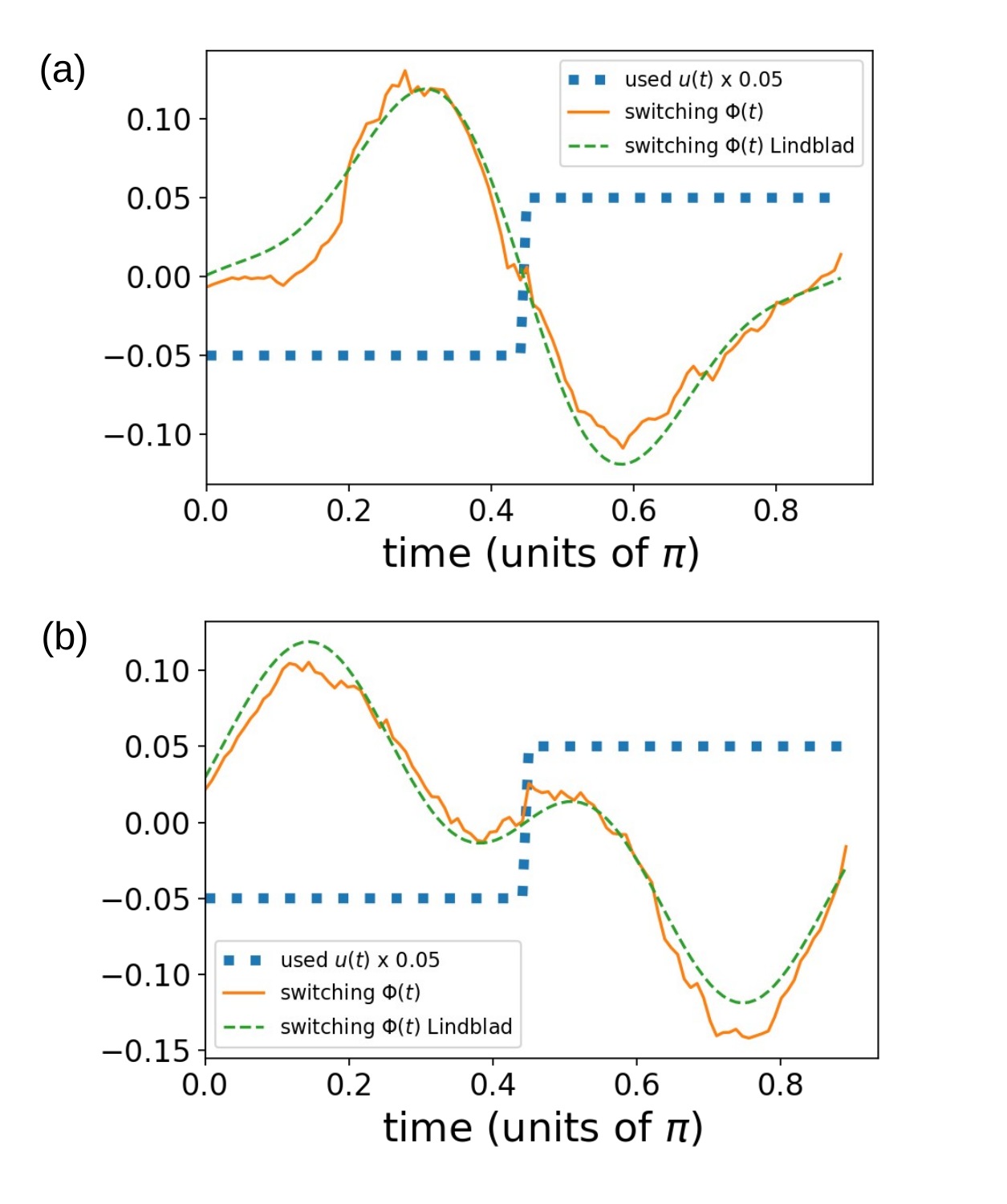}
\caption{The switching function computed using the Lindbladian formalism (Eq.~\eqref{eq:switch}, dashed curves) and using stochastic procedure described by Eq.~\eqref{eqn:quantum_Hoc_switch_2_2} (solid curves, averaging over 500 simulations). Blue dotted curves specify $u(t)$. The initial and target states are: (a) the state retention problem [Eq.~\eqref{eqn:state_retention}]; (b) the state preparation problem [Eq.~\eqref{eqn:state_preparation}].  
}
\label{fig:stochastic_Phi}
\end{center}
\end{figure}

We now compute the switching functions using both procedures outlined in Section \ref{sec:two_procedures}. The first procedure requires the explicit constructions of the density matrix and its costates.  
Using the optimal control for the state retention problem [given in Fig.~\ref{fig:reference}(a)], we compute $\rho$ [Eq.~\eqref{eqn:DM_stochastic}], $\lambda$ [Eq.~\eqref{eqn:lambda_stochastic}], and thus $\Phi$ [Eq.~\eqref{eq:switch}] and $\mathcal{H}_c$ [Eq.~\eqref{eqn:DM_costate_Hoc} and \eqref{eqn:DM_costate_Hoc_diss}]. The results are shown in Fig.~\ref{fig:WF_DM_comparison} and agree well with the exact results (i.e., the results from the deterministic Lindbladian formalism).
Note that in Fig.~\ref{fig:WF_DM_comparison}(c), $\Phi$ and $\mathcal{H}_c$ display larger errors around the switching times; we have tested that this discrepancy becomes weaker upon increasing the time points (not shown). 
To test the second procedure, Fig.~\ref{fig:stochastic_Phi} compares the switching functions computed using Eq.~\eqref{eqn:quantum_Hoc_switch_2_2} 
with the exact ones. In these simulations, the non-optimal control $u(t) = -1 + 2 \Theta(t-t_f/2)$ is used. A good agreement is seen for both state retention and state preparation problems, %
numerically confirming Eq.~\eqref{eqn:quantum_Hoc_switch_2_2}.

\subsection{Optimal control}
\label{sec:example}

\begin{figure}[ht]
\begin{center}
\includegraphics[width=0.8\textwidth]{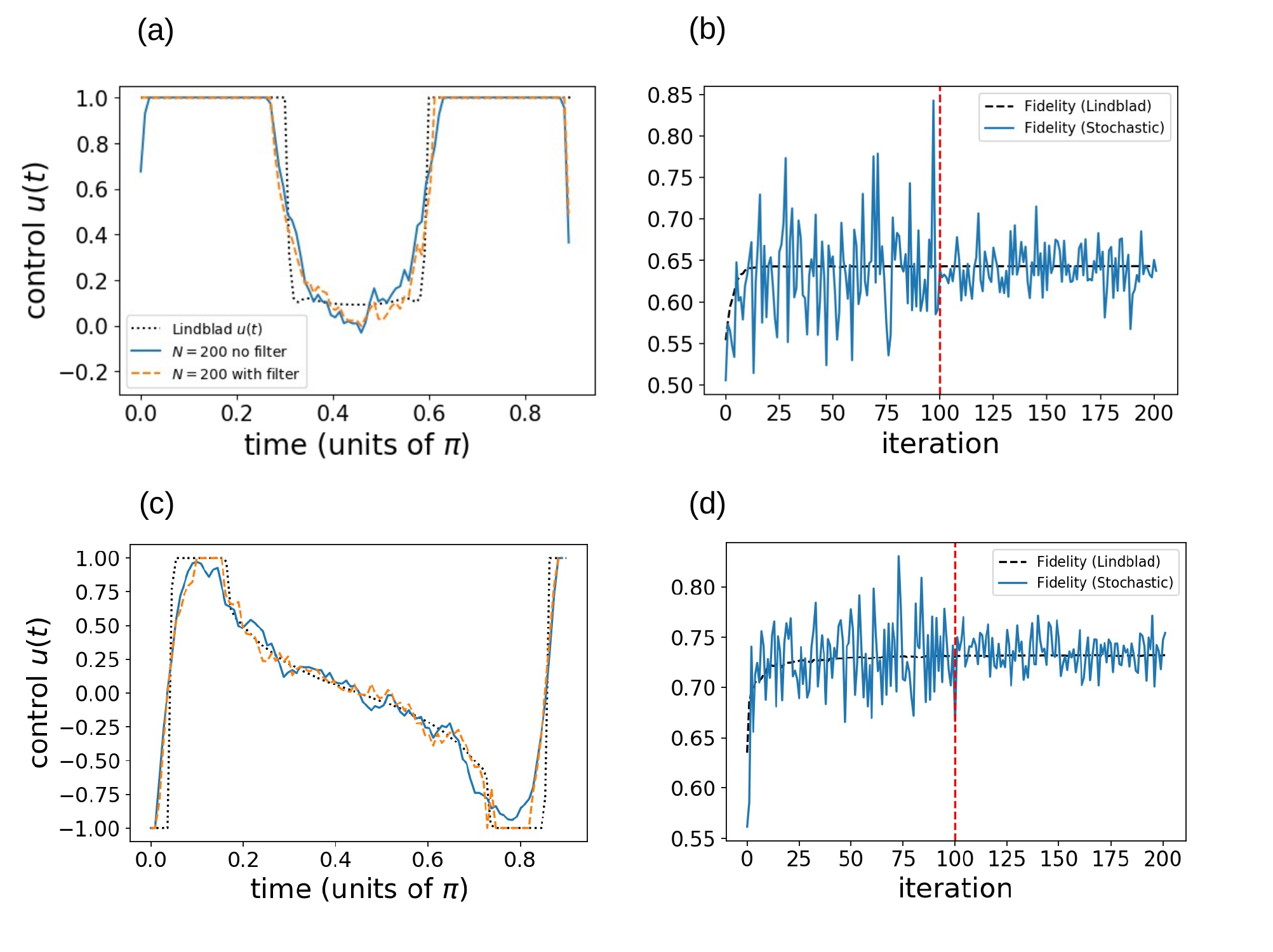}
\caption{ Results of state retention [(a), (b)] and  state preparation [(c), (d)]  problems. 
(a) and (c): Compared to the optimal controls without filtering (solid curves), the optimal controls with the filtering procedure  (Eq.~\eqref{eqn:filter_update}, dashed curves) are closer to those obtained from the Lindbladian formalism (dotted curves). (b) and (d): The fidelities computed from the deterministic Lindbladian (Eq.~\eqref{eq:switch}, dashed curves) and the stochastic Schr\"odinger (Eq.~\eqref{eqn:cost_stoch}, solid curves) formalism. The values agree well. In the stochastic simulations, the first 100 iterations average $50$ realizations whereas the last 100 average $200$ (separated by the red vertical line). The expected fluctuation reduction due to the larger sampling size is clearly seen. 
}
\label{fig:sto_optimal_control}
\end{center}
\end{figure}   

With the switching functions tested, Eq.~\eqref{eqn:quantum_Hoc_switch_2_2} is now applied to determine the optimal control for both state retention and state preparation problems. Due to the noisy nature of stochastic simulations, filtering the functions at some point helps. After some numerical experiments, we adopt two filtering protocols:  $(k+1)$th iteration is updated via
\besub
\begin{align}
\tilde{u}^{(k+1)}(t) &\leftarrow u^{(k)}(t) - \eta \, \tilde{\Phi}^{(k)}(t), 
\label{eqn:TV_filter}\\ 
u^{(k+1)}(t) &\leftarrow P_\varepsilon [ \tilde{u}^{(k+1)}(t) ].
\label{eqn:projection}
\end{align}
\label{eqn:filter_update}
\eesub 
We first elaborate Eq.~\eqref{eqn:TV_filter}.
After $\Phi^{(k)}(t)$ is computed from Eq.~\eqref{eqn:quantum_Hoc_switch_2_2}, it is de-noised by a total-variation (TV) norm \cite{RUDIN1992259}: 
\beq 
\tilde{\Phi}^{(k)}(t) = \underset{y(t)}{ \text{argmin} } \sum_{t_i} \left[
\frac{1}{2} \left[ y(t_i) - \Phi^{(k)}(t_i) \right]^2 + w_{TV} \norm{ y(t_{i+1}) - y(t_i) } \right].
\label{eqn:TV_norm}
\eeq 
$w_{TV} = 0.01$ is adopted and the proximal gradient descent is used to solve Eq.~\eqref{eqn:TV_norm}. We decide to regularize $\Phi(t)$ instead of $u(t)$ because TV norm tends to reduce the magnitude of the discontinuity (jump): there will be a discontinuity in $u(t)$ once $\Phi(t)$ changes sign (the bang-bang protocol) whereas $\Phi(t)$ is generally continuous. 
In Eq.~\eqref{eqn:projection}, the projection $P_\varepsilon$ is defined as 
\beq  
P_\varepsilon[ u(t) ] \equiv 
\begin{cases}
 1 & \text{if } u(t) > 1 - \varepsilon, \\ 
 -1 & \text{if } u(t) < -( 1 - \varepsilon)
\end{cases}
\label{eqn:proj}
\eeq 
$\epsilon = 0.1$ is used. This is designed to promote the bang control around $\Phi(t)=0$. We choose $\varepsilon=0$ in early iterations to avoid biases when  $u(t)$ is still far from the optimal solution. To summarize, three optimization hyperparameters are chosen to be $(\eta, w_{TV}, \varepsilon) = (0.5, 0.01, 0.1)$.

Left panels of Fig.~\ref{fig:sto_optimal_control} shows the optimal control obtained using  Eq.~\eqref{eqn:quantum_Hoc_switch_2_2} for both state retention and  state preparation problems. They generally agree with the exact solutions. 
Compared to the solutions without filtering ($w_{TV}=\varepsilon =0$), the filtering procedure generally results in controls that are smoother over the singular regime (when $\Phi(t)=0$) and sharper near the transition regimes (when $\Phi(t) $ is changing sign or changing between zero and non-zero). Overall the controls with the filtering procedure are closer to the exact solutions. 
Right panels of Fig.~\ref{fig:sto_optimal_control} give the negative of terminal cost function (fidelity) as a function of iterations. The fidelities computed from the deterministic Lindbladian (Eq.~\eqref{eq:switch}, dashed curves) and the stochastic Schr\"odinger (Eq.~\eqref{eqn:cost_stoch}, solid curves) formalism are numerically consistent.  For the stochastic simulation, the first 100 iterations average $N=50$ realizations whereas the last 100 iterations $N=200$. The expected reduction of the fluctuation due to the larger sampling size is also clearly seen. 

\section{Conclusion} 
\label{sec:Conclusion}

A quantum task is mapped to an optimal-control problem once a proper terminal cost function is defined. In terms of optimal control, an optimal solution extremizes the cost function for a given evolution time, and practically the most crucial step to determine the optimal control is the evaluation of the gradient of the terminal cost function. Using PMP, the gradient is efficiently obtained by calculating the switching function that requires solving the original dynamical problem and an auxiliary problem composed of costate variables. In this work, we formulate PMP in terms of density matrices which are essential for open quantum systems and develop a stochastic procedure to compute the switching function for systems obeying Lindbladian dynamics. In particular, we determine the rules to consistently update the wave function and its costate in the stochastic Schr\"odinger formalism so that the switching function can be evaluated without explicit constructing density matrices.  We apply the proposed procedure to determine the optimal control for a dissipative qubit system, and the results are consistent with those obtained directly from the deterministic Lindbladian equation. 
The proposed formalism saves time and memory for simulating large systems, and can be straightforwardly generalized to describe non-Markovian dynamics.

\section*{Acknowledgment}
C.L. thanks Wei-Cheng Lee (Binghamton University, NY, USA) for very helpful discussions.
D.S. acknowledges support from the FWO as post-doctoral fellow of the Research Foundation -- Flanders. 


\bibliography{QC_optimal_control}

\begin{thebibliography}{68}
\expandafter\ifx\csname natexlab\endcsname\relax\def\natexlab#1{#1}\fi
\expandafter\ifx\csname bibnamefont\endcsname\relax
  \def\bibnamefont#1{#1}\fi
\expandafter\ifx\csname bibfnamefont\endcsname\relax
  \def\bibfnamefont#1{#1}\fi
\expandafter\ifx\csname citenamefont\endcsname\relax
  \def\citenamefont#1{#1}\fi
\expandafter\ifx\csname url\endcsname\relax
  \def\url#1{\texttt{#1}}\fi
\expandafter\ifx\csname urlprefix\endcsname\relax\def\urlprefix{URL }\fi
\providecommand{\bibinfo}[2]{#2}
\providecommand{\eprint}[2][]{\url{#2}}

\bibitem[{\citenamefont{Georgescu and Nori}(2012)}]{Georgescu_2012}
\bibinfo{author}{\bibfnamefont{I.}~\bibnamefont{Georgescu}} \bibnamefont{and}
  \bibinfo{author}{\bibfnamefont{F.}~\bibnamefont{Nori}},
  \bibinfo{journal}{Physics World} \textbf{\bibinfo{volume}{25}},
  \bibinfo{pages}{16} (\bibinfo{year}{2012}),
  \urlprefix\url{https://doi.org/10.1088%2F2058-7058%2F25%2F05%2F28}.

\bibitem[{\citenamefont{Nielsen and Chuang}(2011)}]{Neilson_book}
\bibinfo{author}{\bibfnamefont{M.~A.} \bibnamefont{Nielsen}} \bibnamefont{and}
  \bibinfo{author}{\bibfnamefont{I.~L.} \bibnamefont{Chuang}},
  \emph{\bibinfo{title}{Quantum Computation and Quantum Information}}
  (\bibinfo{publisher}{Cambridge University Press, Cambridge},
  \bibinfo{year}{2011}).

\bibitem[{\citenamefont{Kaye et~al.}(2007)\citenamefont{Kaye, Laflamme, and
  Mosca}}]{book:Kaye_book}
\bibinfo{author}{\bibfnamefont{P.}~\bibnamefont{Kaye}},
  \bibinfo{author}{\bibfnamefont{R.}~\bibnamefont{Laflamme}}, \bibnamefont{and}
  \bibinfo{author}{\bibfnamefont{M.}~\bibnamefont{Mosca}},
  \emph{\bibinfo{title}{An introduction to quantum computing}}
  (\bibinfo{publisher}{Oxford University Press, New York},
  \bibinfo{year}{2007}).

\bibitem[{\citenamefont{Shor}(1997)}]{Shor:1997:PAP:264393.264406}
\bibinfo{author}{\bibfnamefont{P.~W.} \bibnamefont{Shor}},
  \bibinfo{journal}{SIAM J. Comput.} \textbf{\bibinfo{volume}{26}},
  \bibinfo{pages}{1484} (\bibinfo{year}{1997}), ISSN \bibinfo{issn}{0097-5397}.

\bibitem[{\citenamefont{Grover}(1996)}]{Grover:1996:FQM:237814.237866}
\bibinfo{author}{\bibfnamefont{L.~K.} \bibnamefont{Grover}}, in
  \emph{\bibinfo{booktitle}{Proceedings of the Twenty-eighth Annual ACM
  Symposium on Theory of Computing}} (\bibinfo{publisher}{ACM, New York},
  \bibinfo{year}{1996}), pp. \bibinfo{pages}{212--219}.

\bibitem[{\citenamefont{Grover}(1997)}]{PhysRevLett.79.325}
\bibinfo{author}{\bibfnamefont{L.~K.} \bibnamefont{Grover}},
  \bibinfo{journal}{Phys. Rev. Lett.} \textbf{\bibinfo{volume}{79}},
  \bibinfo{pages}{325} (\bibinfo{year}{1997}).

\bibitem[{\citenamefont{Peruzzo et~al.}(2014)\citenamefont{Peruzzo, McClean,
  Shadbolt, Yung, Zhou, Love, Aspuru-Guzik, and O'Brien}}]{Peruzzo-2014}
\bibinfo{author}{\bibfnamefont{A.}~\bibnamefont{Peruzzo}},
  \bibinfo{author}{\bibfnamefont{J.}~\bibnamefont{McClean}},
  \bibinfo{author}{\bibfnamefont{P.}~\bibnamefont{Shadbolt}},
  \bibinfo{author}{\bibfnamefont{M.-H.} \bibnamefont{Yung}},
  \bibinfo{author}{\bibfnamefont{X.-Q.} \bibnamefont{Zhou}},
  \bibinfo{author}{\bibfnamefont{P.~J.} \bibnamefont{Love}},
  \bibinfo{author}{\bibfnamefont{A.}~\bibnamefont{Aspuru-Guzik}},
  \bibnamefont{and} \bibinfo{author}{\bibfnamefont{J.~L.}
  \bibnamefont{O'Brien}}, \bibinfo{journal}{Nature Communications}
  \textbf{\bibinfo{volume}{5}}, \bibinfo{pages}{4213} (\bibinfo{year}{2014}).

\bibitem[{\citenamefont{Farhi et~al.}(2014)\citenamefont{Farhi, Goldstone, and
  Gurmann}}]{Farhi_14}
\bibinfo{author}{\bibfnamefont{E.}~\bibnamefont{Farhi}},
  \bibinfo{author}{\bibfnamefont{J.}~\bibnamefont{Goldstone}},
  \bibnamefont{and} \bibinfo{author}{\bibfnamefont{S.}~\bibnamefont{Gurmann}},
  \emph{\bibinfo{title}{A quantum approximate optimization algorithm}}
  (\bibinfo{year}{2014}), \eprint{arXiv:1411.4028}.

\bibitem[{\citenamefont{O'Malley et~al.}(2016)\citenamefont{O'Malley, Babbush,
  Kivlichan, Romero, McClean, Barends, Kelly, Roushan, Tranter, Ding
  et~al.}}]{PhysRevX.6.031007}
\bibinfo{author}{\bibfnamefont{P.~J.~J.} \bibnamefont{O'Malley}},
  \bibinfo{author}{\bibfnamefont{R.}~\bibnamefont{Babbush}},
  \bibinfo{author}{\bibfnamefont{I.~D.} \bibnamefont{Kivlichan}},
  \bibinfo{author}{\bibfnamefont{J.}~\bibnamefont{Romero}},
  \bibinfo{author}{\bibfnamefont{J.~R.} \bibnamefont{McClean}},
  \bibinfo{author}{\bibfnamefont{R.}~\bibnamefont{Barends}},
  \bibinfo{author}{\bibfnamefont{J.}~\bibnamefont{Kelly}},
  \bibinfo{author}{\bibfnamefont{P.}~\bibnamefont{Roushan}},
  \bibinfo{author}{\bibfnamefont{A.}~\bibnamefont{Tranter}},
  \bibinfo{author}{\bibfnamefont{N.}~\bibnamefont{Ding}}, \bibnamefont{et~al.},
  \bibinfo{journal}{Phys. Rev. X} \textbf{\bibinfo{volume}{6}},
  \bibinfo{pages}{031007} (\bibinfo{year}{2016}),
  \urlprefix\url{https://link.aps.org/doi/10.1103/PhysRevX.6.031007}.

\bibitem[{\citenamefont{Preskill}(2018)}]{Preskill2018quantumcomputingin}
\bibinfo{author}{\bibfnamefont{J.}~\bibnamefont{Preskill}},
  \bibinfo{journal}{{Quantum}} \textbf{\bibinfo{volume}{2}},
  \bibinfo{pages}{79} (\bibinfo{year}{2018}), ISSN \bibinfo{issn}{2521-327X},
  \urlprefix\url{https://doi.org/10.22331/q-2018-08-06-79}.

\bibitem[{\citenamefont{Helstrom}(1976)}]{book:Helstrom}
\bibinfo{author}{\bibfnamefont{C.~W.} \bibnamefont{Helstrom}},
  \emph{\bibinfo{title}{Quantum Detection and Estimation Theory}}, Mathematics
  in Science and Engineering 123 (\bibinfo{publisher}{Elsevier, Academic Press,
  New York}, \bibinfo{year}{1976}).

\bibitem[{\citenamefont{Holevo}(2011)}]{book:Holevo}
\bibinfo{author}{\bibfnamefont{A.~S.} \bibnamefont{Holevo}},
  \emph{\bibinfo{title}{Probabilistic and Statistical Aspects of Quantum
  Theory}} (\bibinfo{publisher}{Edizioni della Normale, Superiore Pisa},
  \bibinfo{year}{2011}), \bibinfo{edition}{1st} ed.

\bibitem[{\citenamefont{Giovannetti et~al.}(2006)\citenamefont{Giovannetti,
  Lloyd, and Maccone}}]{PhysRevLett.96.010401}
\bibinfo{author}{\bibfnamefont{V.}~\bibnamefont{Giovannetti}},
  \bibinfo{author}{\bibfnamefont{S.}~\bibnamefont{Lloyd}}, \bibnamefont{and}
  \bibinfo{author}{\bibfnamefont{L.}~\bibnamefont{Maccone}},
  \bibinfo{journal}{Phys. Rev. Lett.} \textbf{\bibinfo{volume}{96}},
  \bibinfo{pages}{010401} (\bibinfo{year}{2006}),
  \urlprefix\url{https://link.aps.org/doi/10.1103/PhysRevLett.96.010401}.

\bibitem[{\citenamefont{Giovannetti et~al.}(2011)\citenamefont{Giovannetti,
  Lloyd, and Maccone}}]{AdvancesQuantumMetrology_2011}
\bibinfo{author}{\bibfnamefont{V.}~\bibnamefont{Giovannetti}},
  \bibinfo{author}{\bibfnamefont{S.}~\bibnamefont{Lloyd}}, \bibnamefont{and}
  \bibinfo{author}{\bibfnamefont{L.}~\bibnamefont{Maccone}},
  \bibinfo{journal}{Nature Photonics} \textbf{\bibinfo{volume}{5}},
  \bibinfo{pages}{222} (\bibinfo{year}{2011}).

\bibitem[{\citenamefont{Tsang et~al.}(2016)\citenamefont{Tsang, Nair, and
  Lu}}]{PhysRevX.6.031033}
\bibinfo{author}{\bibfnamefont{M.}~\bibnamefont{Tsang}},
  \bibinfo{author}{\bibfnamefont{R.}~\bibnamefont{Nair}}, \bibnamefont{and}
  \bibinfo{author}{\bibfnamefont{X.-M.} \bibnamefont{Lu}},
  \bibinfo{journal}{Phys. Rev. X} \textbf{\bibinfo{volume}{6}},
  \bibinfo{pages}{031033} (\bibinfo{year}{2016}),
  \urlprefix\url{https://link.aps.org/doi/10.1103/PhysRevX.6.031033}.

\bibitem[{\citenamefont{Zhuang et~al.}(2017)\citenamefont{Zhuang, Zhang, and
  Shapiro}}]{PhysRevA.96.040304}
\bibinfo{author}{\bibfnamefont{Q.}~\bibnamefont{Zhuang}},
  \bibinfo{author}{\bibfnamefont{Z.}~\bibnamefont{Zhang}}, \bibnamefont{and}
  \bibinfo{author}{\bibfnamefont{J.~H.} \bibnamefont{Shapiro}},
  \bibinfo{journal}{Phys. Rev. A} \textbf{\bibinfo{volume}{96}},
  \bibinfo{pages}{040304} (\bibinfo{year}{2017}),
  \urlprefix\url{https://link.aps.org/doi/10.1103/PhysRevA.96.040304}.

\bibitem[{\citenamefont{Vahlbruch et~al.}(2016)\citenamefont{Vahlbruch, Mehmet,
  Danzmann, and Schnabel}}]{PhysRevLett.117.110801}
\bibinfo{author}{\bibfnamefont{H.}~\bibnamefont{Vahlbruch}},
  \bibinfo{author}{\bibfnamefont{M.}~\bibnamefont{Mehmet}},
  \bibinfo{author}{\bibfnamefont{K.}~\bibnamefont{Danzmann}}, \bibnamefont{and}
  \bibinfo{author}{\bibfnamefont{R.}~\bibnamefont{Schnabel}},
  \bibinfo{journal}{Phys. Rev. Lett.} \textbf{\bibinfo{volume}{117}},
  \bibinfo{pages}{110801} (\bibinfo{year}{2016}),
  \urlprefix\url{https://link.aps.org/doi/10.1103/PhysRevLett.117.110801}.

\bibitem[{\citenamefont{Collaboration}(2011)}]{LIGO_2011}
\bibinfo{author}{\bibfnamefont{T.~L.~S.} \bibnamefont{Collaboration}},
  \bibinfo{journal}{Nature Physics} \textbf{\bibinfo{volume}{7}},
  \bibinfo{pages}{962} (\bibinfo{year}{2011}).

\bibitem[{\citenamefont{Sidhu and Kok}(2020)}]{doi:10.1116/1.5119961}
\bibinfo{author}{\bibfnamefont{J.~S.} \bibnamefont{Sidhu}} \bibnamefont{and}
  \bibinfo{author}{\bibfnamefont{P.}~\bibnamefont{Kok}}, \bibinfo{journal}{AVS
  Quantum Science} \textbf{\bibinfo{volume}{2}}, \bibinfo{pages}{014701}
  (\bibinfo{year}{2020}), \urlprefix\url{https://doi.org/10.1116/1.5119961}.

\bibitem[{\citenamefont{Pezz\`e et~al.}(2018)\citenamefont{Pezz\`e, Smerzi,
  Oberthaler, Schmied, and Treutlein}}]{RevModPhys.90.035005}
\bibinfo{author}{\bibfnamefont{L.}~\bibnamefont{Pezz\`e}},
  \bibinfo{author}{\bibfnamefont{A.}~\bibnamefont{Smerzi}},
  \bibinfo{author}{\bibfnamefont{M.~K.} \bibnamefont{Oberthaler}},
  \bibinfo{author}{\bibfnamefont{R.}~\bibnamefont{Schmied}}, \bibnamefont{and}
  \bibinfo{author}{\bibfnamefont{P.}~\bibnamefont{Treutlein}},
  \bibinfo{journal}{Rev. Mod. Phys.} \textbf{\bibinfo{volume}{90}},
  \bibinfo{pages}{035005} (\bibinfo{year}{2018}),
  \urlprefix\url{https://link.aps.org/doi/10.1103/RevModPhys.90.035005}.

\bibitem[{\citenamefont{Bennett and Wiesner}(1992)}]{PhysRevLett.69.2881}
\bibinfo{author}{\bibfnamefont{C.~H.} \bibnamefont{Bennett}} \bibnamefont{and}
  \bibinfo{author}{\bibfnamefont{S.~J.} \bibnamefont{Wiesner}},
  \bibinfo{journal}{Phys. Rev. Lett.} \textbf{\bibinfo{volume}{69}},
  \bibinfo{pages}{2881} (\bibinfo{year}{1992}),
  \urlprefix\url{https://link.aps.org/doi/10.1103/PhysRevLett.69.2881}.

\bibitem[{\citenamefont{Braunstein and Kimble}(2000)}]{PhysRevA.61.042302}
\bibinfo{author}{\bibfnamefont{S.~L.} \bibnamefont{Braunstein}}
  \bibnamefont{and} \bibinfo{author}{\bibfnamefont{H.~J.}
  \bibnamefont{Kimble}}, \bibinfo{journal}{Phys. Rev. A}
  \textbf{\bibinfo{volume}{61}}, \bibinfo{pages}{042302}
  (\bibinfo{year}{2000}),
  \urlprefix\url{https://link.aps.org/doi/10.1103/PhysRevA.61.042302}.

\bibitem[{\citenamefont{Ekert}(1991)}]{PhysRevLett.67.661}
\bibinfo{author}{\bibfnamefont{A.~K.} \bibnamefont{Ekert}},
  \bibinfo{journal}{Phys. Rev. Lett.} \textbf{\bibinfo{volume}{67}},
  \bibinfo{pages}{661} (\bibinfo{year}{1991}),
  \urlprefix\url{https://link.aps.org/doi/10.1103/PhysRevLett.67.661}.

\bibitem[{\citenamefont{Ursin et~al.}(2007)\citenamefont{Ursin, Tiefenbacher,
  Schmitt-Manderbach, Weier, Scheidl, Lindenthal, Blauensteiner, Jennewein,
  Perdigues, Trojek et~al.}}]{144km}
\bibinfo{author}{\bibfnamefont{R.}~\bibnamefont{Ursin}},
  \bibinfo{author}{\bibfnamefont{F.}~\bibnamefont{Tiefenbacher}},
  \bibinfo{author}{\bibfnamefont{T.}~\bibnamefont{Schmitt-Manderbach}},
  \bibinfo{author}{\bibfnamefont{H.}~\bibnamefont{Weier}},
  \bibinfo{author}{\bibfnamefont{T.}~\bibnamefont{Scheidl}},
  \bibinfo{author}{\bibfnamefont{M.}~\bibnamefont{Lindenthal}},
  \bibinfo{author}{\bibfnamefont{B.}~\bibnamefont{Blauensteiner}},
  \bibinfo{author}{\bibfnamefont{T.}~\bibnamefont{Jennewein}},
  \bibinfo{author}{\bibfnamefont{J.}~\bibnamefont{Perdigues}},
  \bibinfo{author}{\bibfnamefont{P.}~\bibnamefont{Trojek}},
  \bibnamefont{et~al.}, \bibinfo{journal}{Nature Physics}
  \textbf{\bibinfo{volume}{3}}, \bibinfo{pages}{481} (\bibinfo{year}{2007}).

\bibitem[{\citenamefont{Caves and Drummond}(1994)}]{RevModPhys.66.481}
\bibinfo{author}{\bibfnamefont{C.~M.} \bibnamefont{Caves}} \bibnamefont{and}
  \bibinfo{author}{\bibfnamefont{P.~D.} \bibnamefont{Drummond}},
  \bibinfo{journal}{Rev. Mod. Phys.} \textbf{\bibinfo{volume}{66}},
  \bibinfo{pages}{481} (\bibinfo{year}{1994}),
  \urlprefix\url{https://link.aps.org/doi/10.1103/RevModPhys.66.481}.

\bibitem[{\citenamefont{Braunstein and van Loock}(2005)}]{RevModPhys.77.513}
\bibinfo{author}{\bibfnamefont{S.~L.} \bibnamefont{Braunstein}}
  \bibnamefont{and} \bibinfo{author}{\bibfnamefont{P.}~\bibnamefont{van
  Loock}}, \bibinfo{journal}{Rev. Mod. Phys.} \textbf{\bibinfo{volume}{77}},
  \bibinfo{pages}{513} (\bibinfo{year}{2005}),
  \urlprefix\url{https://link.aps.org/doi/10.1103/RevModPhys.77.513}.

\bibitem[{\citenamefont{Weedbrook et~al.}(2012)\citenamefont{Weedbrook,
  Pirandola, Garc\'{\i}a-Patr\'on, Cerf, Ralph, Shapiro, and
  Lloyd}}]{RevModPhys.84.621}
\bibinfo{author}{\bibfnamefont{C.}~\bibnamefont{Weedbrook}},
  \bibinfo{author}{\bibfnamefont{S.}~\bibnamefont{Pirandola}},
  \bibinfo{author}{\bibfnamefont{R.}~\bibnamefont{Garc\'{\i}a-Patr\'on}},
  \bibinfo{author}{\bibfnamefont{N.~J.} \bibnamefont{Cerf}},
  \bibinfo{author}{\bibfnamefont{T.~C.} \bibnamefont{Ralph}},
  \bibinfo{author}{\bibfnamefont{J.~H.} \bibnamefont{Shapiro}},
  \bibnamefont{and} \bibinfo{author}{\bibfnamefont{S.}~\bibnamefont{Lloyd}},
  \bibinfo{journal}{Rev. Mod. Phys.} \textbf{\bibinfo{volume}{84}},
  \bibinfo{pages}{621} (\bibinfo{year}{2012}),
  \urlprefix\url{https://link.aps.org/doi/10.1103/RevModPhys.84.621}.

\bibitem[{\citenamefont{Bao et~al.}(2018)\citenamefont{Bao, Kleer, Wang, and
  Rahmani}}]{PhysRevA.97.062343}
\bibinfo{author}{\bibfnamefont{S.}~\bibnamefont{Bao}},
  \bibinfo{author}{\bibfnamefont{S.}~\bibnamefont{Kleer}},
  \bibinfo{author}{\bibfnamefont{R.}~\bibnamefont{Wang}}, \bibnamefont{and}
  \bibinfo{author}{\bibfnamefont{A.}~\bibnamefont{Rahmani}},
  \bibinfo{journal}{Phys. Rev. A} \textbf{\bibinfo{volume}{97}},
  \bibinfo{pages}{062343} (\bibinfo{year}{2018}).

\bibitem[{\citenamefont{Omran et~al.}(2019)\citenamefont{Omran, Levine,
  Keesling, Semeghini, Wang, Ebadi, Bernien, Zibrov, Pichler, Choi
  et~al.}}]{ahmed19}
\bibinfo{author}{\bibfnamefont{A.}~\bibnamefont{Omran}},
  \bibinfo{author}{\bibfnamefont{H.}~\bibnamefont{Levine}},
  \bibinfo{author}{\bibfnamefont{A.}~\bibnamefont{Keesling}},
  \bibinfo{author}{\bibfnamefont{G.}~\bibnamefont{Semeghini}},
  \bibinfo{author}{\bibfnamefont{T.~T.} \bibnamefont{Wang}},
  \bibinfo{author}{\bibfnamefont{S.}~\bibnamefont{Ebadi}},
  \bibinfo{author}{\bibfnamefont{H.}~\bibnamefont{Bernien}},
  \bibinfo{author}{\bibfnamefont{A.~S.} \bibnamefont{Zibrov}},
  \bibinfo{author}{\bibfnamefont{H.}~\bibnamefont{Pichler}},
  \bibinfo{author}{\bibfnamefont{S.}~\bibnamefont{Choi}}, \bibnamefont{et~al.},
  \bibinfo{journal}{Science} \textbf{\bibinfo{volume}{365}},
  \bibinfo{pages}{570} (\bibinfo{year}{2019}).

\bibitem[{\citenamefont{Friis et~al.}(2018)\citenamefont{Friis, Marty, Maier,
  Hempel, Holz\"apfel, Jurcevic, Plenio, Huber, Roos, Blatt
  et~al.}}]{PhysRevX.8.021012}
\bibinfo{author}{\bibfnamefont{N.}~\bibnamefont{Friis}},
  \bibinfo{author}{\bibfnamefont{O.}~\bibnamefont{Marty}},
  \bibinfo{author}{\bibfnamefont{C.}~\bibnamefont{Maier}},
  \bibinfo{author}{\bibfnamefont{C.}~\bibnamefont{Hempel}},
  \bibinfo{author}{\bibfnamefont{M.}~\bibnamefont{Holz\"apfel}},
  \bibinfo{author}{\bibfnamefont{P.}~\bibnamefont{Jurcevic}},
  \bibinfo{author}{\bibfnamefont{M.~B.} \bibnamefont{Plenio}},
  \bibinfo{author}{\bibfnamefont{M.}~\bibnamefont{Huber}},
  \bibinfo{author}{\bibfnamefont{C.}~\bibnamefont{Roos}},
  \bibinfo{author}{\bibfnamefont{R.}~\bibnamefont{Blatt}},
  \bibnamefont{et~al.}, \bibinfo{journal}{Phys. Rev. X}
  \textbf{\bibinfo{volume}{8}}, \bibinfo{pages}{021012} (\bibinfo{year}{2018}).

\bibitem[{\citenamefont{Doherty et~al.}(2014)\citenamefont{Doherty, Struzhkin,
  Simpson, McGuinness, Meng, Stacey, Karle, Hemley, Manson, Hollenberg
  et~al.}}]{PhysRevLett.112.047601}
\bibinfo{author}{\bibfnamefont{M.~W.} \bibnamefont{Doherty}},
  \bibinfo{author}{\bibfnamefont{V.~V.} \bibnamefont{Struzhkin}},
  \bibinfo{author}{\bibfnamefont{D.~A.} \bibnamefont{Simpson}},
  \bibinfo{author}{\bibfnamefont{L.~P.} \bibnamefont{McGuinness}},
  \bibinfo{author}{\bibfnamefont{Y.}~\bibnamefont{Meng}},
  \bibinfo{author}{\bibfnamefont{A.}~\bibnamefont{Stacey}},
  \bibinfo{author}{\bibfnamefont{T.~J.} \bibnamefont{Karle}},
  \bibinfo{author}{\bibfnamefont{R.~J.} \bibnamefont{Hemley}},
  \bibinfo{author}{\bibfnamefont{N.~B.} \bibnamefont{Manson}},
  \bibinfo{author}{\bibfnamefont{L.~C.~L.} \bibnamefont{Hollenberg}},
  \bibnamefont{et~al.}, \bibinfo{journal}{Phys. Rev. Lett.}
  \textbf{\bibinfo{volume}{112}}, \bibinfo{pages}{047601}
  (\bibinfo{year}{2014}).

\bibitem[{\citenamefont{Farhi et~al.}(2000)\citenamefont{Farhi, Goldstone,
  Gurmann, and Sipser}}]{Farhi_00}
\bibinfo{author}{\bibfnamefont{E.}~\bibnamefont{Farhi}},
  \bibinfo{author}{\bibfnamefont{J.}~\bibnamefont{Goldstone}},
  \bibinfo{author}{\bibfnamefont{S.}~\bibnamefont{Gurmann}}, \bibnamefont{and}
  \bibinfo{author}{\bibfnamefont{M.}~\bibnamefont{Sipser}},
  \emph{\bibinfo{title}{Quantum computation by adiabatic evolution}}
  (\bibinfo{year}{2000}), \eprint{arXiv:quant-ph/0001106}.

\bibitem[{\citenamefont{Rezakhani et~al.}(2009)\citenamefont{Rezakhani, Kuo,
  Hamma, Lidar, and Zanardi}}]{PhysRevLett.103.080502}
\bibinfo{author}{\bibfnamefont{A.~T.} \bibnamefont{Rezakhani}},
  \bibinfo{author}{\bibfnamefont{W.-J.} \bibnamefont{Kuo}},
  \bibinfo{author}{\bibfnamefont{A.}~\bibnamefont{Hamma}},
  \bibinfo{author}{\bibfnamefont{D.~A.} \bibnamefont{Lidar}}, \bibnamefont{and}
  \bibinfo{author}{\bibfnamefont{P.}~\bibnamefont{Zanardi}},
  \bibinfo{journal}{Phys. Rev. Lett.} \textbf{\bibinfo{volume}{103}},
  \bibinfo{pages}{080502} (\bibinfo{year}{2009}),
  \urlprefix\url{https://link.aps.org/doi/10.1103/PhysRevLett.103.080502}.

\bibitem[{\citenamefont{Zhuang}(2014)}]{PhysRevA.90.052317}
\bibinfo{author}{\bibfnamefont{Q.}~\bibnamefont{Zhuang}},
  \bibinfo{journal}{Phys. Rev. A} \textbf{\bibinfo{volume}{90}},
  \bibinfo{pages}{052317} (\bibinfo{year}{2014}),
  \urlprefix\url{https://link.aps.org/doi/10.1103/PhysRevA.90.052317}.

\bibitem[{\citenamefont{Haine and Hope}(2020)}]{PhysRevLett.124.060402}
\bibinfo{author}{\bibfnamefont{S.~A.} \bibnamefont{Haine}} \bibnamefont{and}
  \bibinfo{author}{\bibfnamefont{J.~J.} \bibnamefont{Hope}},
  \bibinfo{journal}{Phys. Rev. Lett.} \textbf{\bibinfo{volume}{124}},
  \bibinfo{pages}{060402} (\bibinfo{year}{2020}),
  \urlprefix\url{https://link.aps.org/doi/10.1103/PhysRevLett.124.060402}.

\bibitem[{\citenamefont{Pontryagin}(1987)}]{book:Pontryagin}
\bibinfo{author}{\bibfnamefont{L.}~\bibnamefont{Pontryagin}},
  \emph{\bibinfo{title}{Mathematical Theory of Optimal Processes}}
  (\bibinfo{publisher}{CRC Press, Boca Raton, FL}, \bibinfo{year}{1987}).

\bibitem[{\citenamefont{Sussmann}(1987)}]{Sussmann_87_01}
\bibinfo{author}{\bibfnamefont{H.~J.} \bibnamefont{Sussmann}},
  \bibinfo{journal}{SIAM Journal on Control and Optimization}
  \textbf{\bibinfo{volume}{25}}, \bibinfo{pages}{433} (\bibinfo{year}{1987}).

\bibitem[{\citenamefont{Luenberger}(1979)}]{book:Luenberger}
\bibinfo{author}{\bibfnamefont{D.~G.} \bibnamefont{Luenberger}},
  \emph{\bibinfo{title}{Introduction to dynamic systems: theory, models, and
  applications}} (\bibinfo{publisher}{Wiley, New York}, \bibinfo{year}{1979}).

\bibitem[{\citenamefont{Heinz~Schattler}(2012)}]{book:GeometricOptimalControl}
\bibinfo{author}{\bibfnamefont{U.~L.} \bibnamefont{Heinz~Schattler}},
  \emph{\bibinfo{title}{Geometric Optimal Control: Theory, Methods and
  Examples}}, Interdisciplinary Applied Mathematics 38
  (\bibinfo{publisher}{Springer-Verlag, New York}, \bibinfo{year}{2012}),
  \bibinfo{edition}{1st} ed., ISBN
  \bibinfo{isbn}{978-1-4614-3833-5,978-1-4614-3834-2}.

\bibitem[{\citenamefont{Lin et~al.}(2020)\citenamefont{Lin, Sels, and
  Wang}}]{PhysRevA.101.022320}
\bibinfo{author}{\bibfnamefont{C.}~\bibnamefont{Lin}},
  \bibinfo{author}{\bibfnamefont{D.}~\bibnamefont{Sels}}, \bibnamefont{and}
  \bibinfo{author}{\bibfnamefont{Y.}~\bibnamefont{Wang}},
  \bibinfo{journal}{Phys. Rev. A} \textbf{\bibinfo{volume}{101}},
  \bibinfo{pages}{022320} (\bibinfo{year}{2020}).

\bibitem[{\citenamefont{Yang et~al.}(2017)\citenamefont{Yang, Rahmani, Shabani,
  Neven, and Chamon}}]{PhysRevX.7.021027}
\bibinfo{author}{\bibfnamefont{Z.-C.} \bibnamefont{Yang}},
  \bibinfo{author}{\bibfnamefont{A.}~\bibnamefont{Rahmani}},
  \bibinfo{author}{\bibfnamefont{A.}~\bibnamefont{Shabani}},
  \bibinfo{author}{\bibfnamefont{H.}~\bibnamefont{Neven}}, \bibnamefont{and}
  \bibinfo{author}{\bibfnamefont{C.}~\bibnamefont{Chamon}},
  \bibinfo{journal}{Phys. Rev. X} \textbf{\bibinfo{volume}{7}},
  \bibinfo{pages}{021027} (\bibinfo{year}{2017}).

\bibitem[{\citenamefont{Lin et~al.}(2019)\citenamefont{Lin, Wang, Kolesov, and
  Kalabic}}]{PhysRevA.100.022327}
\bibinfo{author}{\bibfnamefont{C.}~\bibnamefont{Lin}},
  \bibinfo{author}{\bibfnamefont{Y.}~\bibnamefont{Wang}},
  \bibinfo{author}{\bibfnamefont{G.}~\bibnamefont{Kolesov}}, \bibnamefont{and}
  \bibinfo{author}{\bibfnamefont{U.}~\bibnamefont{Kalabic}},
  \bibinfo{journal}{Phys. Rev. A} \textbf{\bibinfo{volume}{100}},
  \bibinfo{pages}{022327} (\bibinfo{year}{2019}).

\bibitem[{\citenamefont{Zeng and Barnes}(2018)}]{PhysRevA.98.012301}
\bibinfo{author}{\bibfnamefont{J.}~\bibnamefont{Zeng}} \bibnamefont{and}
  \bibinfo{author}{\bibfnamefont{E.}~\bibnamefont{Barnes}},
  \bibinfo{journal}{Phys. Rev. A} \textbf{\bibinfo{volume}{98}},
  \bibinfo{pages}{012301} (\bibinfo{year}{2018}).

\bibitem[{\citenamefont{Hegerfeldt}(2013)}]{PhysRevLett.111.260501}
\bibinfo{author}{\bibfnamefont{G.~C.} \bibnamefont{Hegerfeldt}},
  \bibinfo{journal}{Phys. Rev. Lett.} \textbf{\bibinfo{volume}{111}},
  \bibinfo{pages}{260501} (\bibinfo{year}{2013}).

\bibitem[{\citenamefont{Bukov et~al.}(2018)\citenamefont{Bukov, Day, Sels,
  Weinberg, Polkovnikov, and Mehta}}]{PhysRevX.8.031086}
\bibinfo{author}{\bibfnamefont{M.}~\bibnamefont{Bukov}},
  \bibinfo{author}{\bibfnamefont{A.~G.~R.} \bibnamefont{Day}},
  \bibinfo{author}{\bibfnamefont{D.}~\bibnamefont{Sels}},
  \bibinfo{author}{\bibfnamefont{P.}~\bibnamefont{Weinberg}},
  \bibinfo{author}{\bibfnamefont{A.}~\bibnamefont{Polkovnikov}},
  \bibnamefont{and} \bibinfo{author}{\bibfnamefont{P.}~\bibnamefont{Mehta}},
  \bibinfo{journal}{Phys. Rev. X} \textbf{\bibinfo{volume}{8}},
  \bibinfo{pages}{031086} (\bibinfo{year}{2018}).

\bibitem[{\citenamefont{Sugny et~al.}(2007)\citenamefont{Sugny, Kontz, and
  Jauslin}}]{PhysRevA.76.023419}
\bibinfo{author}{\bibfnamefont{D.}~\bibnamefont{Sugny}},
  \bibinfo{author}{\bibfnamefont{C.}~\bibnamefont{Kontz}}, \bibnamefont{and}
  \bibinfo{author}{\bibfnamefont{H.~R.} \bibnamefont{Jauslin}},
  \bibinfo{journal}{Phys. Rev. A} \textbf{\bibinfo{volume}{76}},
  \bibinfo{pages}{023419} (\bibinfo{year}{2007}),
  \urlprefix\url{https://link.aps.org/doi/10.1103/PhysRevA.76.023419}.

\bibitem[{\citenamefont{Wang et~al.}(2008)\citenamefont{Wang, Huang, and
  Yi}}]{PhysRevA.78.052112}
\bibinfo{author}{\bibfnamefont{L.~C.} \bibnamefont{Wang}},
  \bibinfo{author}{\bibfnamefont{X.~L.} \bibnamefont{Huang}}, \bibnamefont{and}
  \bibinfo{author}{\bibfnamefont{X.~X.} \bibnamefont{Yi}},
  \bibinfo{journal}{Phys. Rev. A} \textbf{\bibinfo{volume}{78}},
  \bibinfo{pages}{052112} (\bibinfo{year}{2008}),
  \urlprefix\url{https://link.aps.org/doi/10.1103/PhysRevA.78.052112}.

\bibitem[{\citenamefont{Stefanatos}(2009)}]{PhysRevA.80.045401}
\bibinfo{author}{\bibfnamefont{D.}~\bibnamefont{Stefanatos}},
  \bibinfo{journal}{Phys. Rev. A} \textbf{\bibinfo{volume}{80}},
  \bibinfo{pages}{045401} (\bibinfo{year}{2009}),
  \urlprefix\url{https://link.aps.org/doi/10.1103/PhysRevA.80.045401}.

\bibitem[{\citenamefont{Ritland and Rahmani}(2018)}]{Ritland_2018}
\bibinfo{author}{\bibfnamefont{K.}~\bibnamefont{Ritland}} \bibnamefont{and}
  \bibinfo{author}{\bibfnamefont{A.}~\bibnamefont{Rahmani}},
  \bibinfo{journal}{New Journal of Physics} \textbf{\bibinfo{volume}{20}},
  \bibinfo{pages}{065005} (\bibinfo{year}{2018}),
  \urlprefix\url{https://doi.org/10.1088%2F1367-2630%2Faaca62}.

\bibitem[{\citenamefont{Deutsch and Jozsa}(1992)}]{Deutsch_Jozsa_92}
\bibinfo{author}{\bibfnamefont{D.}~\bibnamefont{Deutsch}} \bibnamefont{and}
  \bibinfo{author}{\bibfnamefont{R.}~\bibnamefont{Jozsa}},
  \bibinfo{journal}{Proceedings of the Royal Society of London. Series A:
  Mathematical and Physical Sciences} \textbf{\bibinfo{volume}{439}},
  \bibinfo{pages}{553} (\bibinfo{year}{1992}).

\bibitem[{\citenamefont{Simon}(1997)}]{Simon:1997:PQC:264393.264405}
\bibinfo{author}{\bibfnamefont{D.~R.} \bibnamefont{Simon}},
  \bibinfo{journal}{SIAM J. Comput.} \textbf{\bibinfo{volume}{26}},
  \bibinfo{pages}{1474} (\bibinfo{year}{1997}), ISSN \bibinfo{issn}{0097-5397},
  \urlprefix\url{http://dx.doi.org/10.1137/S0097539796298637}.

\bibitem[{\citenamefont{Bennett et~al.}(1997)\citenamefont{Bennett, Bernstein,
  Brassard, and Vazirani}}]{Bennett_97}
\bibinfo{author}{\bibfnamefont{C.}~\bibnamefont{Bennett}},
  \bibinfo{author}{\bibfnamefont{E.}~\bibnamefont{Bernstein}},
  \bibinfo{author}{\bibfnamefont{G.}~\bibnamefont{Brassard}}, \bibnamefont{and}
  \bibinfo{author}{\bibfnamefont{U.}~\bibnamefont{Vazirani}},
  \bibinfo{journal}{SIAM Journal on Computing} \textbf{\bibinfo{volume}{26}},
  \bibinfo{pages}{1510} (\bibinfo{year}{1997}).

\bibitem[{\citenamefont{Harrow et~al.}(2009)\citenamefont{Harrow, Hassidim, and
  Lloyd}}]{PhysRevLett.103.150502}
\bibinfo{author}{\bibfnamefont{A.~W.} \bibnamefont{Harrow}},
  \bibinfo{author}{\bibfnamefont{A.}~\bibnamefont{Hassidim}}, \bibnamefont{and}
  \bibinfo{author}{\bibfnamefont{S.}~\bibnamefont{Lloyd}},
  \bibinfo{journal}{Phys. Rev. Lett.} \textbf{\bibinfo{volume}{103}},
  \bibinfo{pages}{150502} (\bibinfo{year}{2009}),
  \urlprefix\url{https://link.aps.org/doi/10.1103/PhysRevLett.103.150502}.

\bibitem[{\citenamefont{Shor}(1995)}]{PhysRevA.52.R2493}
\bibinfo{author}{\bibfnamefont{P.~W.} \bibnamefont{Shor}},
  \bibinfo{journal}{Phys. Rev. A} \textbf{\bibinfo{volume}{52}},
  \bibinfo{pages}{R2493} (\bibinfo{year}{1995}),
  \urlprefix\url{https://link.aps.org/doi/10.1103/PhysRevA.52.R2493}.

\bibitem[{\citenamefont{Steane}(1996)}]{PhysRevLett.77.793}
\bibinfo{author}{\bibfnamefont{A.~M.} \bibnamefont{Steane}},
  \bibinfo{journal}{Phys. Rev. Lett.} \textbf{\bibinfo{volume}{77}},
  \bibinfo{pages}{793} (\bibinfo{year}{1996}),
  \urlprefix\url{https://link.aps.org/doi/10.1103/PhysRevLett.77.793}.

\bibitem[{\citenamefont{Knill and Laflamme}(1997)}]{PhysRevA.55.900}
\bibinfo{author}{\bibfnamefont{E.}~\bibnamefont{Knill}} \bibnamefont{and}
  \bibinfo{author}{\bibfnamefont{R.}~\bibnamefont{Laflamme}},
  \bibinfo{journal}{Phys. Rev. A} \textbf{\bibinfo{volume}{55}},
  \bibinfo{pages}{900} (\bibinfo{year}{1997}),
  \urlprefix\url{https://link.aps.org/doi/10.1103/PhysRevA.55.900}.

\bibitem[{\citenamefont{Gottesman}(1997)}]{Gottesman-1997}
\bibinfo{author}{\bibfnamefont{D.}~\bibnamefont{Gottesman}},
  \emph{\bibinfo{title}{Stabilizer codes and quantum error correction}}
  (\bibinfo{year}{1997}), \eprint{arXiv:quant-ph/9705052}.

\bibitem[{\citenamefont{Hirose and Cappellaro}(2016)}]{Hirose-2016}
\bibinfo{author}{\bibfnamefont{M.}~\bibnamefont{Hirose}} \bibnamefont{and}
  \bibinfo{author}{\bibfnamefont{P.}~\bibnamefont{Cappellaro}},
  \bibinfo{journal}{Nature} \textbf{\bibinfo{volume}{532}}, \bibinfo{pages}{77}
  (\bibinfo{year}{2016}).

\bibitem[{\citenamefont{Viola and Lloyd}(1998)}]{PhysRevA.58.2733}
\bibinfo{author}{\bibfnamefont{L.}~\bibnamefont{Viola}} \bibnamefont{and}
  \bibinfo{author}{\bibfnamefont{S.}~\bibnamefont{Lloyd}},
  \bibinfo{journal}{Phys. Rev. A} \textbf{\bibinfo{volume}{58}},
  \bibinfo{pages}{2733} (\bibinfo{year}{1998}),
  \urlprefix\url{https://link.aps.org/doi/10.1103/PhysRevA.58.2733}.

\bibitem[{\citenamefont{Viola et~al.}(1999)\citenamefont{Viola, Knill, and
  Lloyd}}]{PhysRevLett.82.2417}
\bibinfo{author}{\bibfnamefont{L.}~\bibnamefont{Viola}},
  \bibinfo{author}{\bibfnamefont{E.}~\bibnamefont{Knill}}, \bibnamefont{and}
  \bibinfo{author}{\bibfnamefont{S.}~\bibnamefont{Lloyd}},
  \bibinfo{journal}{Phys. Rev. Lett.} \textbf{\bibinfo{volume}{82}},
  \bibinfo{pages}{2417} (\bibinfo{year}{1999}),
  \urlprefix\url{https://link.aps.org/doi/10.1103/PhysRevLett.82.2417}.

\bibitem[{\citenamefont{Benedetti et~al.}(2013)\citenamefont{Benedetti,
  Buscemi, Bordone, and Paris}}]{PhysRevA.87.052328}
\bibinfo{author}{\bibfnamefont{C.}~\bibnamefont{Benedetti}},
  \bibinfo{author}{\bibfnamefont{F.}~\bibnamefont{Buscemi}},
  \bibinfo{author}{\bibfnamefont{P.}~\bibnamefont{Bordone}}, \bibnamefont{and}
  \bibinfo{author}{\bibfnamefont{M.~G.~A.} \bibnamefont{Paris}},
  \bibinfo{journal}{Phys. Rev. A} \textbf{\bibinfo{volume}{87}},
  \bibinfo{pages}{052328} (\bibinfo{year}{2013}),
  \urlprefix\url{https://link.aps.org/doi/10.1103/PhysRevA.87.052328}.

\bibitem[{\citenamefont{Uhlmann}(1976)}]{UHLMANN1976273}
\bibinfo{author}{\bibfnamefont{A.}~\bibnamefont{Uhlmann}},
  \bibinfo{journal}{Reports on Mathematical Physics}
  \textbf{\bibinfo{volume}{9}}, \bibinfo{pages}{273 } (\bibinfo{year}{1976}),
  ISSN \bibinfo{issn}{0034-4877},
  \urlprefix\url{http://www.sciencedirect.com/science/article/pii/0034487776900604}.

\bibitem[{\citenamefont{Jozsa}(1994)}]{doi:10.1080/09500349414552171}
\bibinfo{author}{\bibfnamefont{R.}~\bibnamefont{Jozsa}},
  \bibinfo{journal}{Journal of Modern Optics} \textbf{\bibinfo{volume}{41}},
  \bibinfo{pages}{2315} (\bibinfo{year}{1994}).

\bibitem[{Hc_()}]{Hc_basic2}
\bibinfo{note}{The control Hamiltonian is defined as the inner product of a
  costate $\lambda$ and the time-derivative of the state $\dot{x}$. The space
  of matrices forms a Hilbert space, and adopting the Hilbert-Schmidt inner
  product immediately leads to Eq.~\eqref{eqn:Hc_basic}. Alst note that when
  $\rho$ and $\lambda$ are Hermitian and the dynamics is a proper completely
  positive and trace-preserving map, Tr$[ \lambda \dot{\rho}]$ is guaranteed to
  be real.}

\bibitem[{\citenamefont{Breuer and Petruccione}(2002)}]{book:open_quantum}
\bibinfo{author}{\bibfnamefont{H.-P.} \bibnamefont{Breuer}} \bibnamefont{and}
  \bibinfo{author}{\bibfnamefont{F.}~\bibnamefont{Petruccione}},
  \emph{\bibinfo{title}{The Theory of Open Quantum Systems}}
  (\bibinfo{publisher}{Oxford University Press}, \bibinfo{year}{2002}).

\bibitem[{\citenamefont{Dalibard et~al.}(1992)\citenamefont{Dalibard, Castin,
  and M\o{}lmer}}]{PhysRevLett.68.580}
\bibinfo{author}{\bibfnamefont{J.}~\bibnamefont{Dalibard}},
  \bibinfo{author}{\bibfnamefont{Y.}~\bibnamefont{Castin}}, \bibnamefont{and}
  \bibinfo{author}{\bibfnamefont{K.}~\bibnamefont{M\o{}lmer}},
  \bibinfo{journal}{Phys. Rev. Lett.} \textbf{\bibinfo{volume}{68}},
  \bibinfo{pages}{580} (\bibinfo{year}{1992}),
  \urlprefix\url{https://link.aps.org/doi/10.1103/PhysRevLett.68.580}.

\bibitem[{\citenamefont{Castin et~al.}(2008)\citenamefont{Castin, Dalibard, and
  Molmer}}]{castin2008wave}
\bibinfo{author}{\bibfnamefont{Y.}~\bibnamefont{Castin}},
  \bibinfo{author}{\bibfnamefont{J.}~\bibnamefont{Dalibard}}, \bibnamefont{and}
  \bibinfo{author}{\bibfnamefont{K.}~\bibnamefont{Molmer}},
  \emph{\bibinfo{title}{A wave function approach to dissipative processes}}
  (\bibinfo{year}{2008}), \eprint{0805.4002}.

\bibitem[{\citenamefont{Rudin et~al.}(1992)\citenamefont{Rudin, Osher, and
  Fatemi}}]{RUDIN1992259}
\bibinfo{author}{\bibfnamefont{L.~I.} \bibnamefont{Rudin}},
  \bibinfo{author}{\bibfnamefont{S.}~\bibnamefont{Osher}}, \bibnamefont{and}
  \bibinfo{author}{\bibfnamefont{E.}~\bibnamefont{Fatemi}},
  \bibinfo{journal}{Physica D: Nonlinear Phenomena}
  \textbf{\bibinfo{volume}{60}}, \bibinfo{pages}{259 } (\bibinfo{year}{1992}),
  ISSN \bibinfo{issn}{0167-2789},
  \urlprefix\url{http://www.sciencedirect.com/science/article/pii/016727899290242F}.

\end{thebibliography}

\appendix

\end{document}